\documentclass[aps,pre,preprint,onecolumn,citeautoscript,superscriptaddress]{revtex4-1}  
\usepackage{amsmath,amssymb,bm} 
\usepackage{graphicx}  
\usepackage[tight]{subfigure}    
\usepackage{color} 
\usepackage{placeins}
\usepackage[papersize={8.5in,11in}]{geometry}
\usepackage[colorlinks=true]{hyperref}  
\hypersetup{
    bookmarks=true,         
    unicode=false,          
    pdftoolbar=true,        
    pdfmenubar=true,        
    pdffitwindow=false,     
    pdfstartview={FitH},    
    pdftitle={My title},    
    pdfauthor={Author},     
    pdfsubject={Subject},   
    pdfcreator={Creator},   
    pdfproducer={Producer}, 
    pdfkeywords={keyword1} {key2} {key3}, 
    pdfnewwindow=true,      
    colorlinks=true,       
    linkcolor=magenta, 
    citecolor=blue,        
    filecolor=magenta,      
    urlcolor=cyan           
} 

\newcommand{\beq}{\begin{equation}}
\newcommand{\eeq}{\end{equation}}
\newcommand{\nn}{\nonumber \\}
\def\bea{\begin{eqnarray}}
\def\eea{\end{eqnarray}}

\begin{document}
\title{Numerical study of fermion and boson models\\ with infinite-range random interactions}  
 \author{Wenbo Fu}
 \affiliation{Department of Physics, Harvard University, Cambridge, Massachusetts, 02138, USA}
 \author{Subir Sachdev}
 \affiliation{Department of Physics, Harvard University, Cambridge, Massachusetts, 02138, USA}
 \affiliation{Perimeter Institute for Theoretical Physics, Waterloo, Ontario N2L 2Y5, Canada}
 \date{\today \\
 \vspace{0.6in}}
\begin{abstract}
We present numerical studies of fermion and boson models with random all-to-all interactions (the SYK models).
The high temperature expansion and exact diagonalization of the $N$-site fermion model are used to compute
the entropy density: our results are consistent with the numerical solution of $N=\infty$ saddle point equations,
and the presence of a non-zero entropy density in the limit of vanishing temperature. The exact diagonalization results for the fermion Green's
function also appear to converge well to the $N=\infty$ solution. For the hard-core boson model, the exact diagonalization study indicates
spin glass order. Some results on the entanglement entropy and the out-of-time-order correlators are also presented.
\end{abstract}
\maketitle 
\section{Introduction}
\label{sec:intro}

Fermion and boson models with infinite-range random interactions 
were studied in the 1990's and later \cite{sachdev1992gapless,PG98,GPS99,georges2001quantum,MJR02,MJR03}
as models of quantum systems with novel non-Fermi liquid or spin glass ground states.
More recently, it was proposed that such models are holographically connected to the dynamics of AdS$_2$ horizons of 
charged black holes \cite{SS10,SS10b}, and remarkable connections have since emerged to topics in 
quantum chaos and 
black hole physics \cite{kitaev2015talk,sachdev2015Bh,maldacena2015abound,hosur2015chaos,JPRV16,YLX16,Denef16,Jevicki16}. 

The model introduced by Sachdev and Ye \cite{sachdev1992gapless} was defined on $N$ sites, and each site had particles with $M$ flavors; 
then the double limit $N\rightarrow \infty$,
followed by $M \rightarrow \infty$, was taken. In such a limit, the random interactions depend on 2 indices, each taking $N$ values.
Taking the double limit is challenging in numerical studies, and so they have been restricted to $M=2$ with increasing values of $N$ \cite{MJR02,MJR03}.
It was found that the ground state for $N \rightarrow \infty$ at a fixed $M=2$ was almost certainly a spin glass.
So a direct numerical test of the more exotic non-Fermi liquid states has not so far been possible.

Kitaev \cite{kitaev2015talk} has recently introduced an alternative large $N$ limit in which the random interaction depends upon
4 indices, each taking $N$ values; the saddle-point equations in the $N\rightarrow \infty$ limit are the same as those in 
Ref.~\onlinecite{sachdev1992gapless}.
No separate $M \rightarrow\infty$ is required, and this is a significant advantage for numerical study. 
The present paper will study such
Sachdev-Ye-Kitaev (SYK) models by exact diagonalization; some additional results will also be obtained in a high temperature
expansion. Our numerical studies will be consistent the fermionic SYK model displaying a non-Fermi liquid state which has
extensive entropy, and entanglement entropy, in the zero temperature limit. 
For the case of the bosonic SYK model, our numerical study indicates spin-glass order: this implies that the analytic study
of the large $N$ limit will require replica symmetry breaking \cite{GPS99}.

The outline of this paper is as follows. In Section~\ref{sec:largeN}, we review the large $N$ solution of the SYK model,
and present new results on its high temperature expansion.
In Section~\ref{sec:EDf} we present exact diagonalization results for the fermionic SYK model, while the hard-core boson case is 
considered later in Section~\ref{sec:EDb}. Section~\ref{sec:OTOC} contains a few results on out-of-time-order correlators of recent interest.

\section{Large $N$ limit for fermions}
\label{sec:largeN}

This section will introduce the SYK model for complex fermions, and review its large $N$ limit.
We will obtain expressions for the fermion Green's function and the free energy density.
A high temperature expansion for these quantities will appear in Section~\ref{sec:hte}.

The Hamiltonian of the SYK model is
\beq
H = \frac{1}{(2 N)^{3/2}} \sum_{i,j,k,\ell=1}^N J_{ij;k\ell} \, c_i^\dagger c_j^\dagger c_k^{\vphantom \dagger} c_\ell^{\vphantom \dagger} 
- \mu \sum_{i} c_i^\dagger c_i^{\vphantom \dagger} \label{H}
\eeq
where the $J_{ij;k\ell}$ are complex Gaussian random couplings with zero mean obeying
\bea
J_{ji;k\ell} = - J_{ij;k\ell} \quad , \quad
J_{ij;\ell k} &=& - J_{ij;k\ell} \quad , \quad
J_{k\ell;ij} = J_{ij;k\ell}^\ast \nn
\overline{|J_{ij;k\ell}|^2} &=& J^2.
\eea
The above Hamiltonian can be viewed as a `matrix model'
on Fock space, with a dimension which is exponential in $N$. But notice that there are only order $N^4$ independent matrix elements,
and so Fock space matrix elements are highly correlated.
The conserved U(1) density, $\mathcal{Q}$ is related to the average fermion number by
\beq
\mathcal{Q} = \frac{1}{N} \sum_i \left\langle c_i^\dagger c_i^{\vphantom \dagger} \right\rangle. \label{defQ}
\eeq
The value of $\mathcal{Q}$ can be varied by the chemical potential $\mu$, and ranges between 0 an 1. The solution described below
applies for the range of $\mu$ for which $0 < \mathcal{Q} < 1$, and so realizes a compressible state. 

Using the imaginary-time path-integral formalism, the partition function can be written as
\beq
\mathcal{Z}=\int Dc^\dagger Dc \exp{(-\mathcal{S})}
\eeq
where
\beq
\mathcal{S}=\int_0^{\beta}d\tau(c^\dagger\partial_{\tau}c+H),
\eeq
where $\beta = 1/T$ is the inverse temperature, and we have already changed the operator $c$ into a Grassman number. 

In the replica trick, we take $n$ replicas of the system and then take the $n\rightarrow 0$ limit
\beq
\ln{\mathcal{Z}}=\lim_{n\rightarrow 0}\frac{1}{n}(\mathcal{Z}^n-1)
\eeq
Introducing replicas $c_{ia}$, with $a=1 \ldots n$, we can average over disorder and obtain the replicated imaginary time ($\tau$) action
\beq
\mathcal{S}_n =\sum_{ia}  \int_0^{\beta} d \tau c_{ia}^\dagger \left( \frac{\partial}{\partial \tau} - \mu \right) c_{ia} 
- \frac{J^2}{4N^3} \sum_{ab} \int_0^{\beta} d \tau d \tau'
\left| \sum_i c_{ia}^\dagger (\tau) c_{i b}^{\vphantom\dagger} (\tau') \right|^4 ; \label{Sn}
\eeq
(here we neglect normal-ordering corrections which vanish as $N \rightarrow \infty$). Then the partition function can be written as
\beq
\mathcal{Z}^n=\int\prod_{a}Dc_a^\dagger D c_a \exp{(-\mathcal{S}_n)}
\eeq

Notice that the action has a global SU($N$) symmetry under $c_{i a} \rightarrow U_{ij} 
c_{j a}$. Also,
if we ignore the time-derivative term in Eq.~(\ref{Sn}), notice that the action has a 
U(1) gauge invariance under $c_{i a} \rightarrow e^{i \vartheta_i (\tau)}
c_{j a}$. And indeed, in the low energy limit leading to Eq.~(\ref{Gz}), the time-derivative term can be neglected. However, we cannot
drop the time-derivative term at the present early stage, as it plays a role in selecting the manner in which the U(1) 
gauge invariance is `broken' in 
the low energy limit. In passing, we note that this phenomenon appears to be analogous to that described in the holographic study of non-Fermi 
liquids by DeWolfe {\em et al.} \cite{DGR11}: there, the bulk fermion representing the low energy theory is also argued to acquire the color
degeneracy of the boundary fermions due to an almost broken gauge invariance. As in Ref.~\cite{DGR11}, we expect the bulk degrees of freedom
of gravitational duals to the SYK model to carry a density of order $N$ \cite{sachdev2015Bh}.

Following the earlier derivation \cite{sachdev1992gapless}, we decouple the interaction by two successive Hubbard-Stratonovich transformations.
First, we introduce the real field $Q_{ab} (\tau, \tau')$ obeying
\beq
Q_{ab} (\tau, \tau') = Q_{ba} (\tau', \tau).
\eeq
The equation above is required because the action is invariant under the reparameterization $a\leftrightarrow b, \tau\leftrightarrow \tau'$.
In terms of this field
\bea
\mathcal{S}_n &=& \sum_{ia}  \int_0^{\beta} d \tau c_{ia}^\dagger \left( \frac{\partial}{\partial \tau} - \mu \right) c_{ia} 
+ \sum_{ab} \int_0^{\beta} d \tau d \tau' \Biggl\{ \frac{N}{4J^2} \left[ Q_{ab} (\tau, \tau') \right]^2 \nn
&~&~~~~~~~~~~~~~~~~~ - \frac{1}{2N}
Q_{ab} (\tau, \tau') \left| \sum_i c_{ia}^\dagger (\tau) c_{i b}^{\vphantom\dagger} (\tau') \right|^2 \Biggr\}.
\eea
A second decoupling with the complex field $P_{ab} (\tau, \tau') $ obeying
\beq
P_{ab} (\tau, \tau') = P_{ba}^\ast (\tau', \tau)
\eeq
yields
\bea
\mathcal{S}_n &=& \sum_{ia}  \int_0^{\beta} d \tau c_{ia}^\dagger \left( \frac{\partial}{\partial \tau} - \mu \right) c_{ia} 
+ \sum_{ab} \int_0^{\beta} d \tau d \tau' \Biggl\{ \frac{N}{4J^2} \left[ Q_{ab} (\tau, \tau') \right]^2  +
\frac{N}{2} Q_{ab} (\tau, \tau') \left|P_{ab} (\tau, \tau')\right|^2 
\nn 
&~&~~~~~~~~~~~~~~~~~~~~ -  Q_{ab} (\tau, \tau') P_{ba} (\tau', \tau)
 \sum_i c_{ia}^\dagger (\tau) c_{i b}^{\vphantom\dagger} (\tau')  \Biggr\}
\eea

\begin{figure}[!htb]
\center
\includegraphics[width=4.75in]{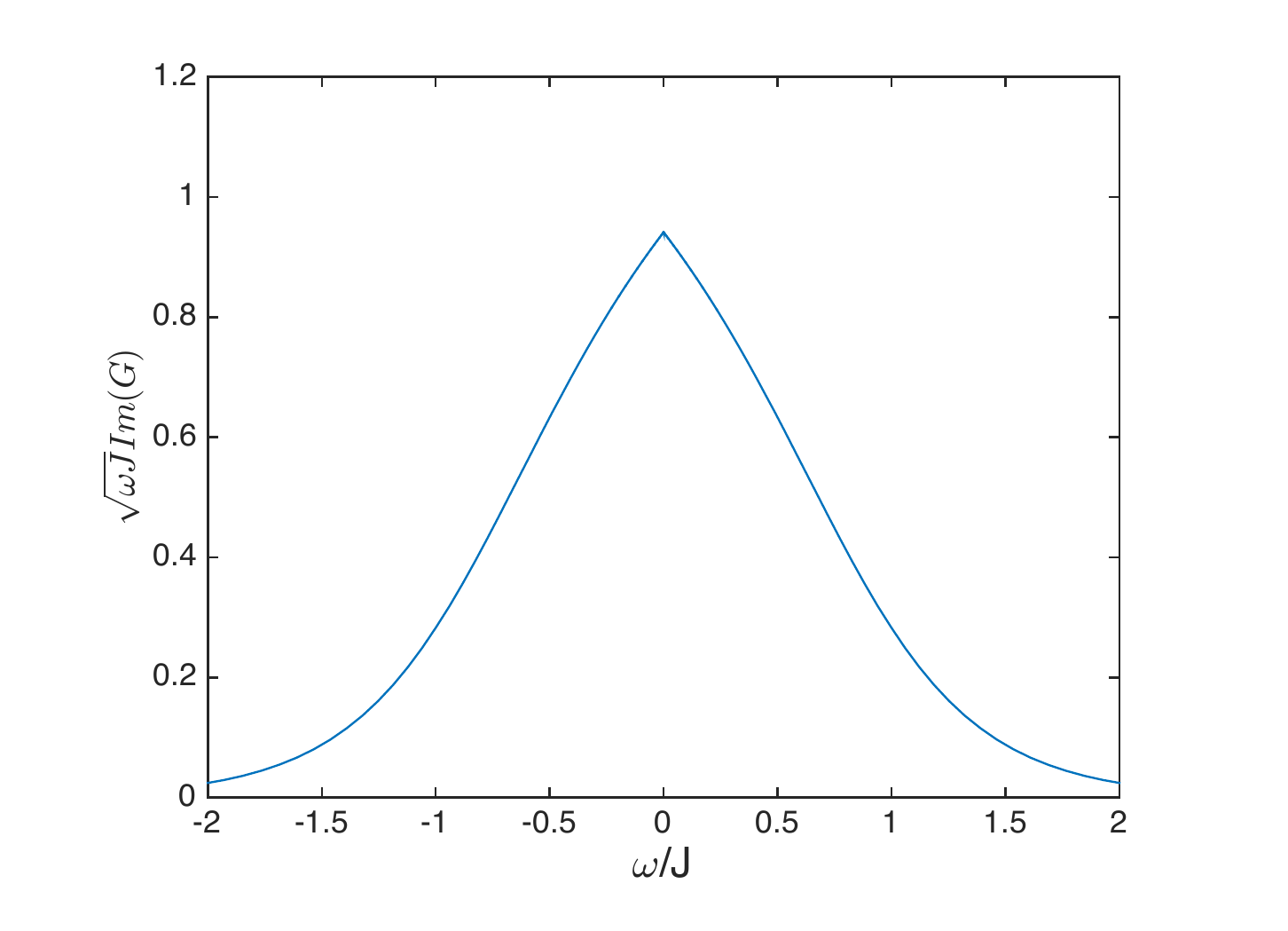}
\caption{Figure, adapted from Ref~\onlinecite{sachdev1992gapless}, showing the imaginary part of Green's function multiplied by $\sqrt{\omega}$ as a function of $\omega$ at particle-hole symmetric point $\theta=0$. Our definition of the
Green's function, Eq.~(\ref{Green function}), differs by a sign from Ref.~\cite{sachdev1992gapless}.  \label{SY}}
\end{figure}

Now we study the saddle point of this action in the large $N$ limit.  After integrating out fermion field and take $\frac{\delta S}{\delta P_{ba}}=0$, we obtain
\beq
P_{ab}(\tau,\tau')=\frac{1}{N}\langle c_{ia}^\dagger(\tau)c_{ib}(\tau')\rangle
\eeq
Note that we have combined $\frac{N}{2}Q_{ab}|P_{ab}|^2$ and $\frac{N}{2}Q_{ba}|P_{ba}|^2$ as one term.  
Similarly, taking derivative with respect to $Q_{ab}$, we have
\beq
Q_{ab}(\tau,\tau')=J^2|P_{ab}(\tau,\tau')|^2.
\eeq
If we only consider diagonal solution in the replica space (non spin-glass state), we can define the self energy:
\beq
\Sigma(\tau,\tau')=-Q(\tau,\tau')P(\tau',\tau),
\eeq
and the Green's function
\beq
G(\tau,\tau')=-\langle T_{\tau}c(\tau)c^\dagger(\tau')\rangle.\label{Green function}
\eeq
Then we have
\beq
P(\tau,\tau')=G(\tau',\tau),
\eeq
and the saddle point solution becomes
\bea
G(i\omega_n)=\frac{1}{i\omega_n+\mu-\Sigma(i\omega_n)} \nn
\Sigma(\tau)=-J^2 G^2(\tau) G(-\tau)\label{EOM}
\eea
The above equation shows a re-parameterization symmetry at low temperature if we ignore the 
$i\omega_n$ term \cite{kitaev2015talk,sachdev2015Bh}. At zero temperature, the low energy Green's function is found to be\cite{sachdev1992gapless,sachdev2015Bh}
\beq
G(z)=C\frac{e^{-i(\pi/4+\theta)}}{\sqrt{z}}\quad,\quad \mbox{Im}(z)>0,~|z|\ll J,~T=0 \label{Gz}
\eeq
where $C$ is a positive number, and $-\pi/4<\theta<\pi/4$ characterizes the particle-hole asymmetry. 
A full numerical solution for Eq.~(\ref{EOM}) at zero temperature was also obtained in Ref.~\onlinecite{sachdev1992gapless}, and is shown in Fig.~\ref{SY}. 
We can see the $1/\sqrt{z}$ behavior at low energy.
However, it is not possible to work entirely within this low energy scaling limit to obtain other low temperature properties: the $i \omega_n$ 
term is needed to properly regularize the ultraviolet, and select among the many possible solutions of the low-energy 
equations \cite{georges2001quantum,sachdev2015Bh}.

\subsection{Free energy and thermal entropy}
The free energy is defined to be
\beq
\mathcal{F}=-\frac{1}{\beta}\ln{\mathcal{Z}_{eff}}
\eeq
where $\mathcal{Z}_{eff}$ has only one replica. So
\beq
\mathcal{Z}_{eff}=\int Dc^\dagger D c \exp{(-\mathcal{S})},
\eeq
with
\bea
\mathcal{S} &=& \sum_{i}  \int_0^{\beta} d \tau c_{i}^\dagger \left( \frac{\partial}{\partial \tau} - \mu \right) c_{i} 
+ \int_0^{\beta} d \tau d \tau' \Biggl\{ \frac{N}{4J^2} \left[ Q(\tau, \tau') \right]^2  +
\frac{N}{2} Q (\tau, \tau') \left|P(\tau, \tau')\right|^2 
\nn 
&~&~~~~~~~~~~~~~~~~~~~~ -  Q (\tau, \tau') P(\tau', \tau)
 \sum_i c_i^\dagger (\tau) c_i^{\vphantom\dagger} (\tau')  \Biggr\}
\eea
For the free energy density $\mathcal{F}/N$, we can just drop the site index $i$ to give the single site action, substituting the Green's function and self energy
\bea
\mathcal{S} &=&  \int_0^{\beta} d \tau d\tau' c^\dagger(\tau) \left( \frac{\partial}{\partial \tau}\delta(\tau-\tau') - \mu\delta(\tau-\tau') +\Sigma(\tau,\tau')\right) c(\tau') \nn 
&~&~~~~~~
+ \int_0^{\beta} d \tau d \tau' \Biggl\{ \frac{1}{4J^2} \left[ \Sigma(\tau,\tau')/G(\tau,\tau') \right]^2  -\frac{1}{2}\Sigma(\tau,\tau')G(\tau',\tau)
 \Biggr\}
\eea
After integrating out the fermion field
\beq
\mathcal{S} =  -\operatorname{Tr} \ln{\left[(\partial_{\tau}-\mu)\delta(\tau-\tau')+\Sigma(\tau,\tau')\right]}
+ \int_0^{\beta} d \tau d \tau' \Biggl\{ \frac{1}{4J^2} \left[ \Sigma(\tau,\tau')/G(\tau,\tau') \right]^2  -\frac{1}{2}\Sigma(\tau,\tau')G(\tau',\tau)
 \Biggr\}
\eeq
To verify this result, we can vary with respect to $\Sigma(\tau,\tau')$ and $G(\tau,\tau')$, also using the fact that $\Sigma(\tau,\tau')=\Sigma^*(\tau',\tau)$, $G(\tau,\tau')=G^*(\tau',\tau)$, to obtain the equations of motions as before.

In the large $N$ limit, we can substitute in the classical solution, and then free energy density is 
\beq
\frac{\mathcal{F}}{N}=T \sum_n \ln{(-\beta G(i\omega_n))}-\int_0^{\beta} d \tau \frac{3}{4}\Sigma(\tau)G(-\tau)
\eeq
The thermal entropy density can be obtained by
\beq
\frac{S}{N} =-\frac{1}{N}\frac{\partial \mathcal{F}}{\partial T}
\eeq

\subsection{High temperature expansion}
\label{sec:hte}

Now we present a solution of Eqs.~(\ref{EOM}) by a high temperature expansion (HTE).
Equivalently, this can be viewed as an expansion in powers of $J$.

We will limit ourselves to the simpler particle-hole symmetric case, $\mathcal{Q}=1/2$, for which both $G$ and $\Sigma$
are odd functions of $\omega_n$. We start with the high temperature limit
\beq
G_0 (i \omega_n) = \frac{1}{i \omega_n}
\eeq
and then expand both $G$ and $\Sigma$ in powers of $J^2$: $G=G_0+G_1+\cdots$ and $\Sigma=\Sigma_0+\Sigma_1+\cdots$.
The successive terms can be easily obtained by iteratively expanding both equations in Eq.~(\ref{EOM}), and repeatedly performing
Fourier transforms between frequency and time.
\bea
\Sigma_1(i\omega_n)&=&J^2\frac{1}{4i\omega_n}\nn
G_1(i\omega_n)&=&J^2\frac{1}{4(i\omega_n)^3}\nn
\Sigma_2(i\omega_n)&=&J^4\frac{3}{16(i\omega_n)^3}\nn
G_2(i\omega_n)&=&J^4\frac{1}{4(i\omega_n)^5}\nn
\Sigma_3(i\omega_n)&=&J^6\left[\frac{15}{32(i\omega_n)^5}+\frac{3}{128 T^2 (i\omega_n)^3}\right]\nn
G_3(i\omega_n)&=&J^6\left[\frac{37}{64(i\omega_n)^7}+\frac{3}{128 T^2(i\omega_n)^5}\right]\nn
\Sigma_4(i\omega_n)&=&J^8\left[\frac{561}{256 (i\omega_n)^7}+\frac{75}{512 T^4 (i\omega_n)^5}-\frac{1}{256 T^4(i\omega_n)^3}\right]\nn
G_4(i\omega_n)&=&J^8\left[\frac{5}{2(i\omega_n)^9}+\frac{81}{512T^2(i\omega_n)^7}-\frac{1}{256T^4(i\omega_n)^5}\right]
\eea

The free energy density can be written in terms of $G(i\omega_n)$ and $\Sigma(i\omega_n)$
\beq
\frac{\mathcal{F}}{N}=T \sum_n \ln{(-\beta G(i\omega_n))}-\frac{3T}{4}\sum_n \Sigma(i\omega_n)G(i\omega_n)
\eeq

We also need to regularize the above free energy by subtracting and adding back the free particle part
\beq
\frac{\mathcal{F}}{N}=-T\ln{2}+T \sum_n \left[\ln{(-\beta G(i\omega_n))}-\ln{(-\beta i\omega_n)}\right]-\frac{3T}{4}\sum_n \Sigma(i\omega_n)G(i\omega_n)
\eeq

The series expansion of the entropy density is
\beq
\frac{\mathcal{S}}{N}=\ln{2}-\frac{1}{64}\frac{J^2}{T^2}+\frac{1}{512}\frac{J^4}{T^4}-\frac{11}{36864}\frac{J^6}{T^6}+\frac{599}{11796480}\frac{J^8}{T^8}+\cdots\label{HTE entropy}
\eeq

\begin{figure}[!htb]
\center
\includegraphics[width=4.75in]{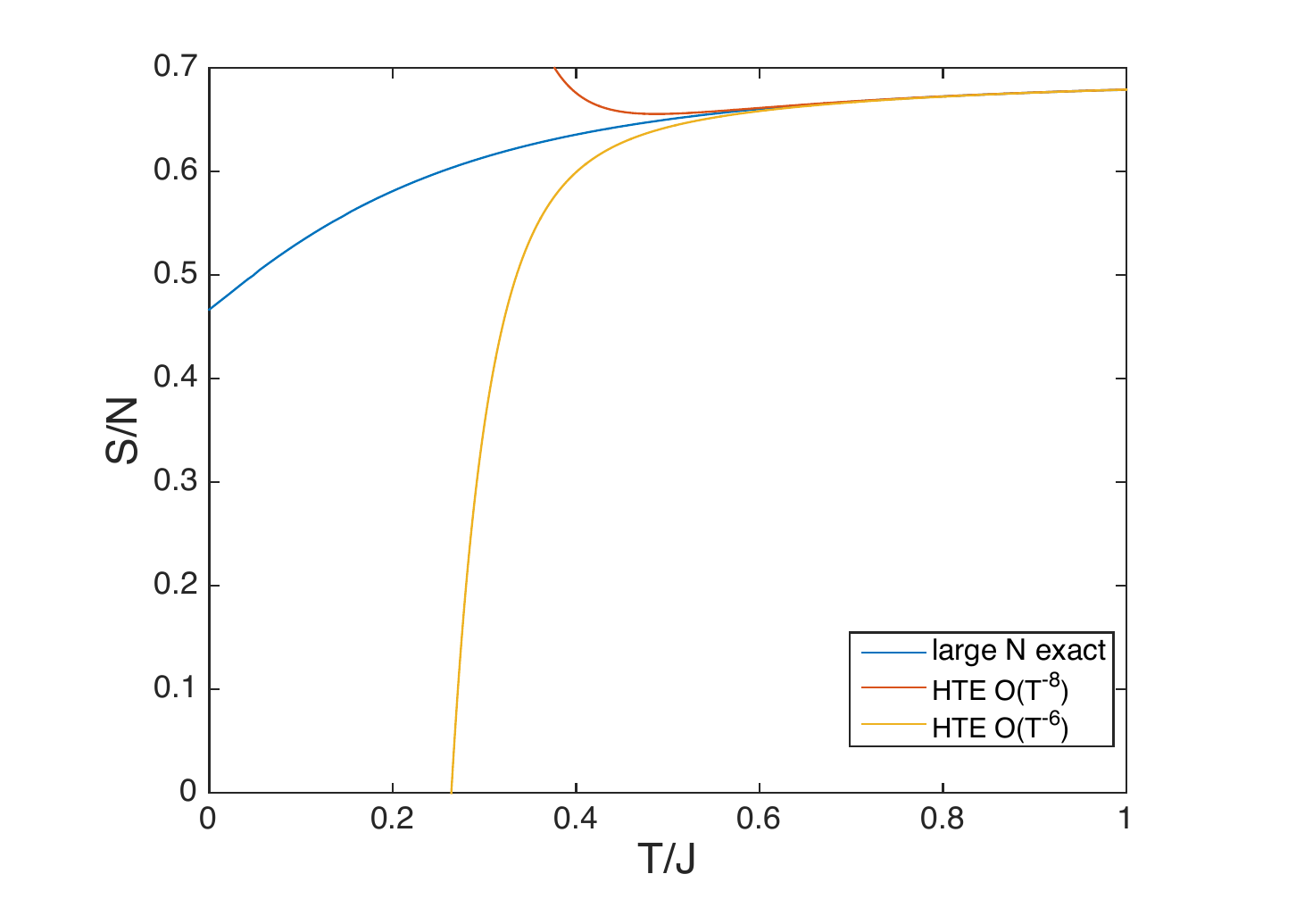}
\caption{Entropy computation from exact large $N$ EOM and HTE: at hight temperature, all approaches the infinite temperature limit $S/N=\ln 2$. HTE result fit the exact result quite well for $T/J>0.6$.}. 
\label{HTEcompare}
\end{figure}

Next, we describe our numerical solution of Eq.~(\ref{EOM}) at non-zero temperature. 
We used a Fourier transform (FT) to iterate between the two equations, until we obtained a convergent solution. For faster convergence, 
we started at high temperature, and used the above high temperature expansion as the initial form. Then we decreased temperature to get the full temperature dependence. We compare the large $N$ exact numerical result with the high temperature expansion
in Fig.~\ref{HTEcompare}. At high temperatures, all methods converge to $\ln{2}$ as expected. The HTE results fit the exact numerics 
quite well for $T/J>0.6$, but are no longer accurate at lower $T$. The exact numerics shows a finite entropy density in the limit of 
vanishing temperature, with a value consistent with earlier analytic results \cite{georges2001quantum}\cite{sachdev2015Bh}.

\FloatBarrier

\section{Exact diagonalization for fermions}
\label{sec:EDf}

We now test the validity of the large $N$ results by comparing with an exact diagonalization (ED) computation
at finite $N$. For the numerical setup, it was useful to employ the Jordan-Wigner transformation to map the Hamiltonian to a spin model
\beq
c_{i}=\sigma_{i}^-\prod_{j<i}\sigma_j^z, \quad c^\dagger_{i}=\sigma_{i}^+\prod_{j<i}\sigma_j^z
\eeq
We built a matrix of the $N$ spins and diagonalized it numerically. After obtaining the full spectrum, we obtained both the imaginary part of Green's function and thermal entropy, and our results are compared with the large $N$ results in Fig.~\ref{SY} and Fig.~\ref{HTEcompare}.

In this note, we focus on the particle-hole symmetric point. But particle-hole symmetry does not correspond to the point $\mu=0$ in the Hamiltonian Eq.~(\ref{H}), because there are quantum corrections to the chemical potential $\delta \mu\sim\mathcal{O}(N^{-1})$ coming from the terms in
which $i,j,k,l$ are not all different from each other, because these terms are not particle-hole symmetric. So we use a Hamiltonian with extra correction terms that compensate $\delta \mu$:
\beq
H = \frac{1}{(2 N)^{3/2}} \sum_{i,j,k,\ell=1}^N J_{ij;k\ell} \, (c_i^\dagger c_j^\dagger c_k^{\vphantom \dagger} c_\ell^{\vphantom \dagger}+\delta_{ik}n \, c^\dagger_jc_l-\delta_{il}n \, c^\dagger_j c_k-\delta_{jk}n\, c^\dagger_ic_l+\delta_{jl}n\, c^\dagger_ic_k),
\label{symmetricH}
\eeq
where we use $n=1/2$ for the particle-hole symmetric case.

We define the on-site retarded Green's function by
\beq
G^R_i(t,t')=-i\theta(t-t')\langle \{c_i(t),c^\dagger_i(t')\} \rangle.
\eeq
Using Lehmann representation
\beq
G^R_i(\omega)=\frac{1}{Z}\sum_{n n'}\frac{\langle n|c_i|n'\rangle\langle n'|c_i^\dagger|n\rangle}{\omega+E_n-E_{n'}+i\eta}(e^{-\beta E_n}+e^{-\beta E_{n'}}),
\eeq
at zero temperature, we obtain
\beq
G^R_i(\omega)=\sum_{n'}\frac{\langle 0|c_i|n'\rangle \langle n'|c_i^\dagger|0\rangle}{\omega+E_0-E_{n'}+i\eta}+\frac{\langle 0|c_i^\dagger|n'\rangle \langle n'|c_i|0\rangle}{\omega-E_0+E_{n'}+i\eta}.
\eeq
Using $(\omega+i\eta)^{-1}=\mathcal{P}\frac{1}{\omega}-i\pi\delta(\omega)$
\beq
\text{Im}G^R_i(\omega)=-\pi\sum_{n'}\left[\langle 0|c_i|n'\rangle \langle n'|c_i^\dagger|0\rangle\delta(\omega+E_0-E_{n'})+\langle 0|c_i^\dagger|n'\rangle \langle n'|c_i|0\rangle\delta(\omega-E_0+E_{n'})\right].\label{rho}
\eeq
Numerically, we replace the delta function with a Lorentzian by taking a small $\eta$:
\beq
\delta(E_0-E_{n'}+\omega)=\lim_{\eta\rightarrow 0^+}\frac{1}{\pi}\frac{\eta}{(E_0-E_{n'}+\omega)^2+\eta^2}
\eeq

A subtlety in the above numerics, when $\mu=0$, is the presence of an anti-unitary particle-hole symmetry. The ground state turns out to be doubly degenerate for some system sizes. If so, we will have two ground states $|0\rangle$ and $|0'\rangle$ in the expression of $G_i^R(\omega)$, and we need to sum them up to get the correct Green's function. 

To better understand this degeneracy, we can define the particle-hole transformation operator
\beq
P=\prod_i(c_i^\dagger+c_i) K\label{PH}
\eeq
where $K$ is the anti-unitary operator. One can show that it is a symmetry of our Hamiltonian Eq.~(\ref{symmetricH}), $\left[H,P\right]=0$. When the total site number $N$ is odd, we know any eigenstate $|\Psi\rangle$ and its particle-hole partner $P|\Psi\rangle$ must be different and degenerate. For even site number, these two states may be the same state. However for $N=2 \mod{4}$, one can show that $P^2=-1$, and then the degeneracy is analogous to the time-reversal Kramers doublet for $T^2=-1$ particles. We expect that all the eigenvalues must be doubly degenerate. For $N=0\mod{4}$, $P^2=1$, there is no protected degeneracy in the half filing sector. These facts were all checked by numerics, and carefully considered in the calculation of the Green's function. 

A better understanding of the above facts can be reached from the perspective of symmetry-protected topological (SPT) phases.
As shown recently in Ref.~\onlinecite{YLX16}, the complex SYK model can be thought of as the boundary of a 1D SPT system in the 
symmetry class AIII. The periodicity of 4 in $N$ arises from the fact that we need to put 4 chains to gap out the boundary 
degeneracy without breaking the particle-hole symmetry. In the Majorana SYK case, the symmetric Hamiltonian 
can be constructed as a symmetric matrix in the Clifford algebra $\textit{Cl}_{0,N-1}$, and the Bott periodicity in the 
real representation of the Clifford algebra gives rise to a $\mathbb{Z}_8$ classification\cite{YLX16}. Here, for the complex SYK case, we can similarly construct the Clifford algebra by dividing one complex fermion into two Majorana fermions, and then we will have a periodicity of 4.

\subsection{Green's function}

From the above definition of retarded Green's function, we can relate them to the imaginary time Green's function as defined in Eq.~(\ref{Green function}), $G^R(\omega)=G(i\omega_n\rightarrow\omega+i\eta)$. 
\begin{figure}[!htb]
\center
\includegraphics[width=4.75in]{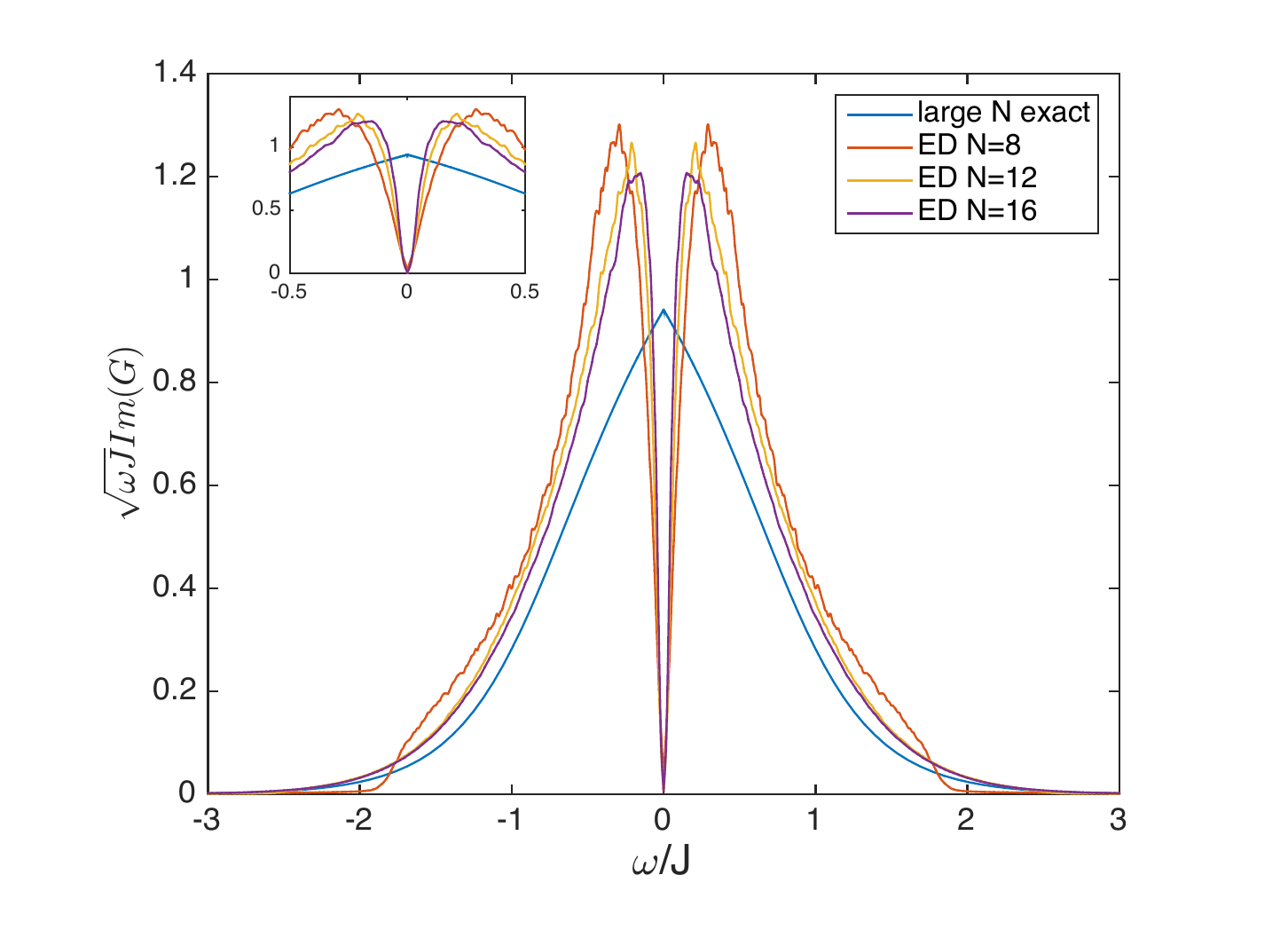}
\caption{Imaginary part of  the Green's function in real frequency space from large $N$ and exact diagonalization. The inset figure is zoomed in near $\omega=0$.}. \label{G*sqrt(w)}
\end{figure}
In Fig.~\ref{G*sqrt(w)}, we show a comparison between the imaginary part of the Green's function from large $N$, 
and from the exact diagonalization computation.  The spectral function from ED is particle-hole symmetric for all $N$, this is guaranteed by the particle-hole symmetry and can be easily shown from the definition of the spectral function Eq.~(\ref{rho}). The two results agree well
at high frequencies. At low frequencies, the deviations between the exact diagonalization and large $N$ results get smaller at larger $N$.

\begin{figure}[!htb]
\center
\includegraphics[width=4.75in]{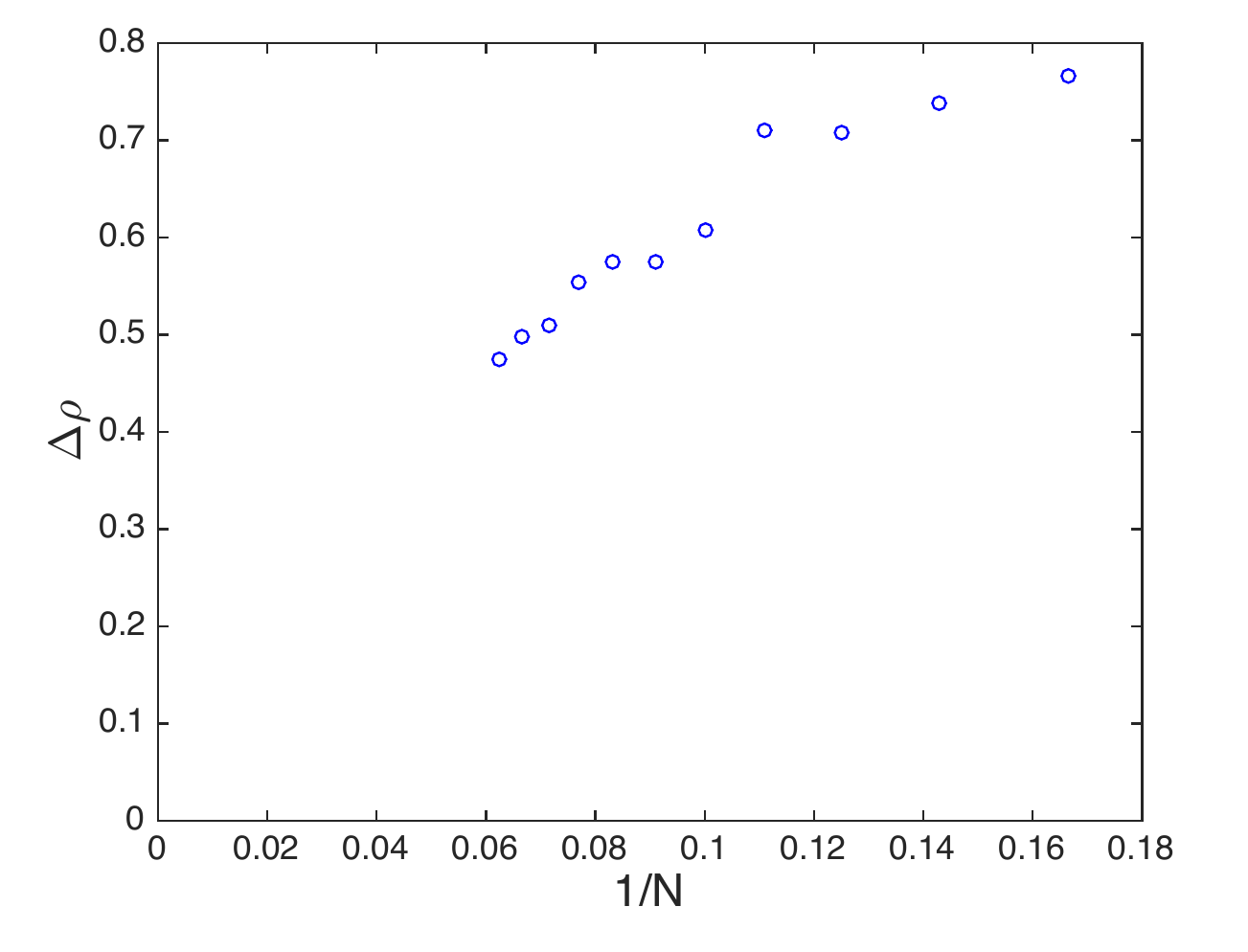}
\caption{The difference of integrated spectral function between ED at different N and large N result. The difference appears to be tending to 0 as 
$N$ approaches infinity.} 
\label{scaling}
\end{figure}

For a quantitative estimate of the deviations between the large $N$ and exact diagonalization results, we
we compute the areas under each curve in Fig.~\ref{G*sqrt(w)}, and compare their difference:
\beq
\Delta \rho=\int d\omega |\text{Im}G_{ED}(\omega)-\text{Im}G_{N=\infty}(\omega)|
\eeq
As shown in Fig.~\ref{scaling}, the convergence to the $N=\infty$ limit is slow, possibly with a power smaller than $1/N$.

\FloatBarrier

\subsection{Entropy}

\begin{figure}[!htb]
\center
\includegraphics[width=4.75in]{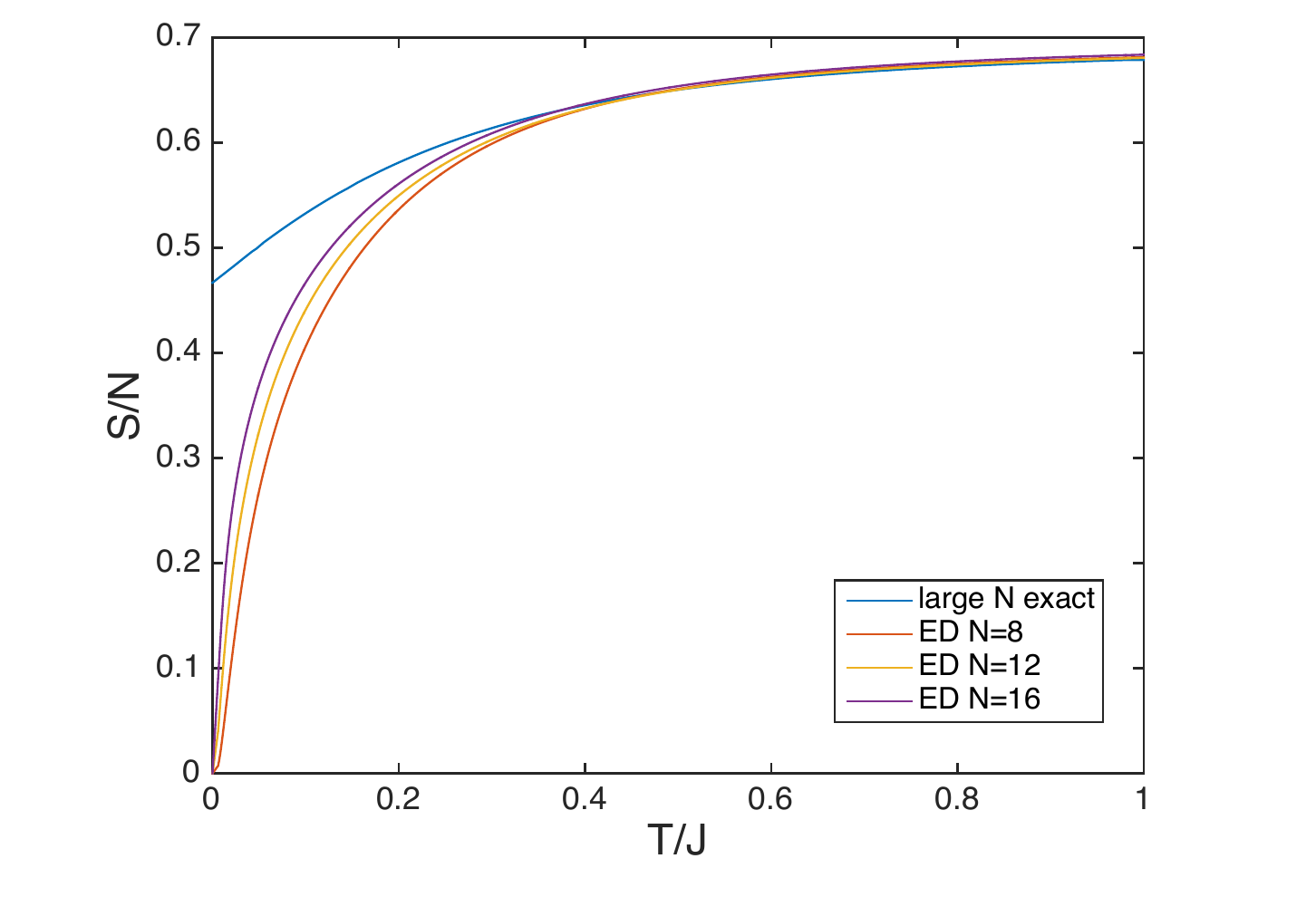}
\caption{Thermal entropy computation from ED, and large $N$. At high temperature, all results the infinite temperature limit $S/N=\ln 2$. At low temperature, all ED results go to zero, but do approach the $N = \infty$ results with increasing $N$. Note that the limits $N \rightarrow \infty$
and $T \rightarrow 0$ do not commute, and the non-zero entropy as $T \rightarrow 0$ is obtained only when the $N\rightarrow \infty$ is taken first}. 
\label{EDcompare}
\end{figure}
\begin{figure}[!htb]
\center
\includegraphics[width=4.75in]{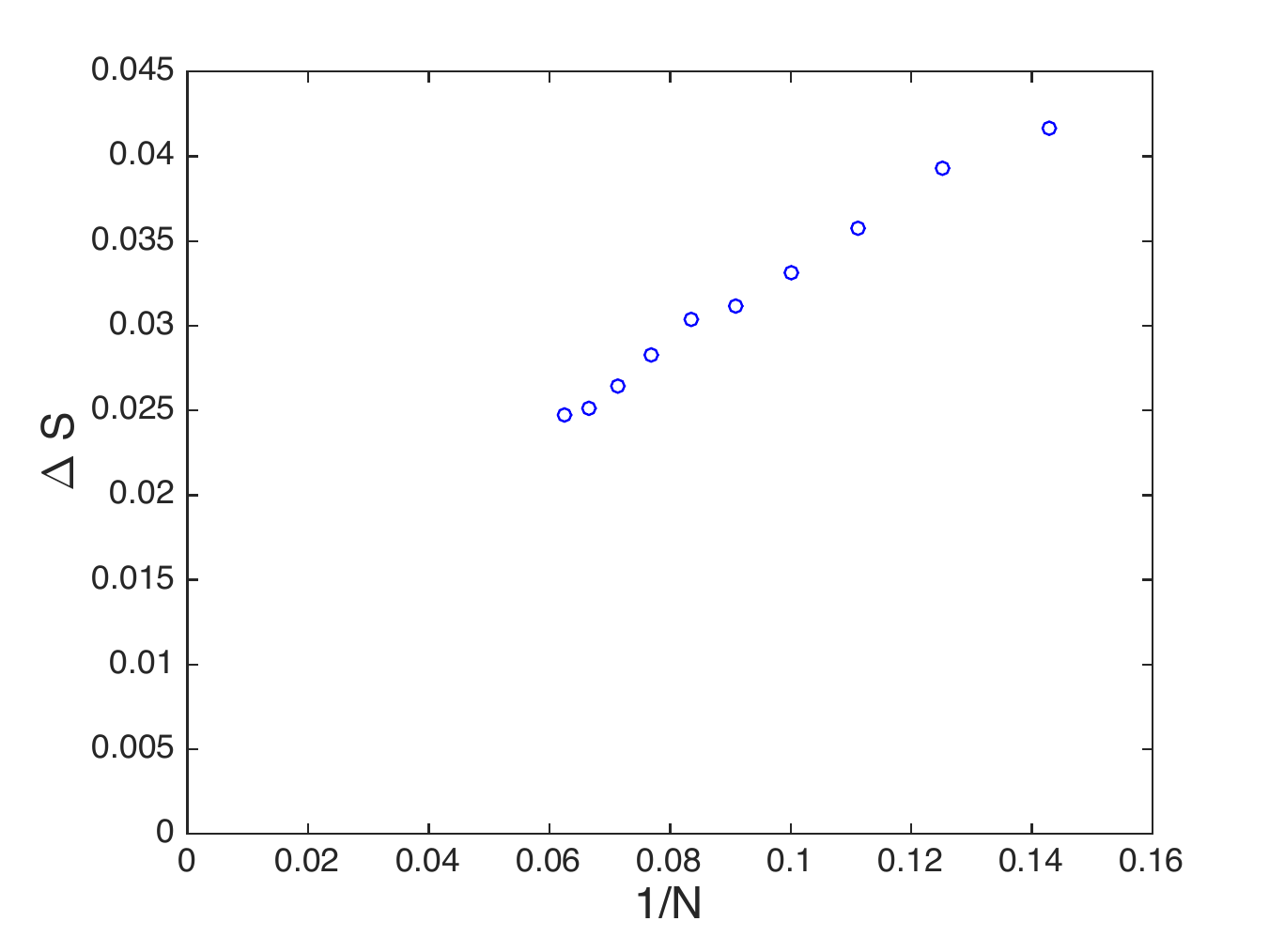}
\caption{The difference of integrated thermal entropy between ED at different N and large N result. The difference goes to 0 as $1/N$ approaches 0.}. 
\label{Sscaling}
\end{figure}
We can also compute the finite temperature entropy from ED. The partition function can be obtained from the full spectrum
\beq
\mathcal{Z}=\sum_n e^{-\beta E_n},
\eeq
where $E_n$ is the many-body energy, and then free energy density is 
\beq
\frac{F}{N}=-\frac{\beta}{N}\log{\mathcal{Z}}.
\eeq
We can obtain the entropy density from
\beq
\frac{S}{N}=\frac{1}{N}\frac{\langle E\rangle - F}{ T}
\eeq
where $\langle E\rangle=\sum_n\frac{E_n e^{-\beta E_n}}{\mathcal{Z}}$ is the average energy.

We use this approach to compute the thermal entropy from the full spectrum, and compare it with the thermal entropy calculated from large $N$ equations of motion Eq.~(\ref{EOM}). As shown in Fig.~\ref{EDcompare}, the 
finite size ED computation gives rise to the correct limit $s=\ln 2$ in the high temperature regime, and it agrees with the large $N$ result quite well for  $T/J>0.5$. Although there is a clear trend that a larger system size gives rise to larger thermal entropy at low temperature, we cannot obtain a finite zero temperature entropy for any finite $N$. This is due to the fact that the non-zero zero temperature entropy is obtained by taking the large $N$ limit first then taking the zero temperature limit.

As in Fig.~\ref{scaling}, we estimate the deviation from the large $N$ theory by defining
\beq
\Delta S=\int dT |S_{ED}(T)/N-S_{N=\infty}(T)/N_{\infty}|,
\eeq
and plot the result in Fig.~\ref{Sscaling}.
The finite size correction goes to 0 as $1/N$ goes to zero.

\FloatBarrier

\subsection{Entanglement entropy}

\begin{figure}[!htb]
\center
\includegraphics[width=4.75in]{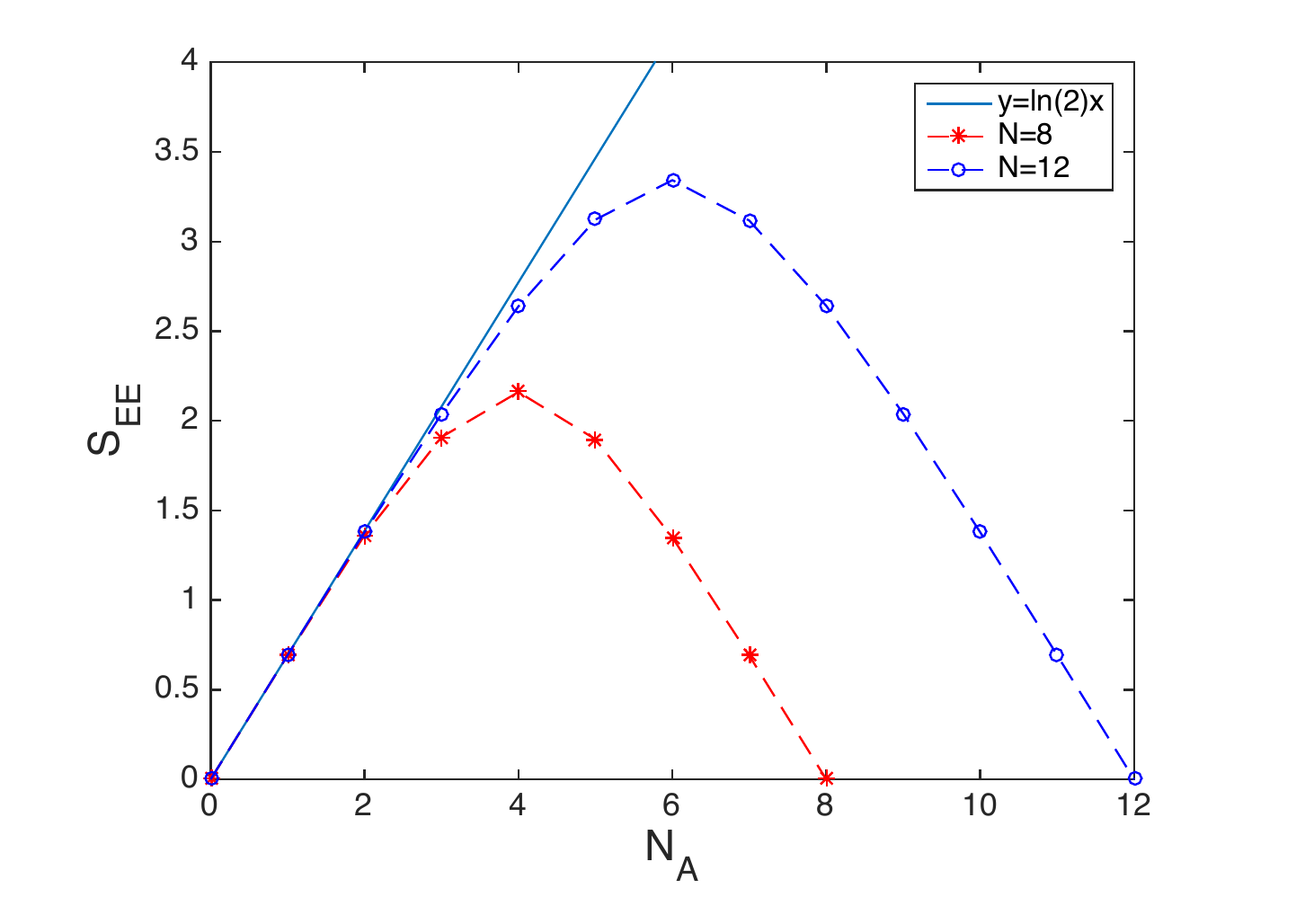}
\caption{Entanglement entropy for the ground state. We divide the system into two subsystems, A and B, we trace out part B and 
calculate the entropy for the reduced density matrix $\rho_A$. The x-axis is the size of subsystem A.} 
\label{EE}
\end{figure}
Finally, we compute the entanglement entropy in the ground state, obtained by choosing a subsystem A of $N$ sites, and tracing over
the remaining sites; the results are in Fig.~\ref{EE}. For $N_A < N/2$, we find that $S_{EE}$ is proportional to $N_A$, thus obeying the volume
law, and so even the ground state obeys eigenstate thermalization \cite{YLX16}. 
We would expect that $S_{EE}/N_A$ equals the zero temperature limit of the entropy density 
$S/N$ \cite{tarun}. However, our value of $S_{EE}/N_A$ appears closer to $\ln 2$
(see Fig.~\ref{EE}) than the value of $S/N$ as $T \rightarrow 0$. Given the small difference between $\ln 2 = 0.69$ and $S/N (T \rightarrow 0) = 0.464848...$,
we expect this is a finite-size discrepancy.

\FloatBarrier

\section{Out-of-time-order correlations and scrambling}
\label{sec:OTOC}

One of the interesting properties of the SYK model is that it exhibits quantum chaos \cite{kitaev2015talk,JPRV16}. The quantum chaos can be quantified in terms of an out-of-time-ordered correlator $\langle A(t)B(0)A(t)B(0)\rangle$ (OTOC) obtained from the cross terms in $\langle \left[A(t),B(0)\right]^2\rangle$ \cite{maldacena2015abound}. 
The exponential decay in the OTOC results in an exponential growth of $\langle \left[A(t),B(0)\right]^2\rangle$ at short times,
and the latter was connected to analogous behavior in classical chaos.
In particular, Ref.~\onlinecite{maldacena2015abound}, established a rigorous bound, $2 \pi/\beta$, 
for the decay rate, $\lambda_L$, 
of the OTOC, and the Majorana SYK model  is expected \cite{kitaev2015talk} to saturate this bound in the strong-coupling 
limit $\beta J\gg1$. In the opposite perturbative limit, $\beta J\ll 1$, one expects $\lambda_L\sim J$. 
Ref.~\onlinecite{hosur2015chaos} performed a ED calculation of the OTOC on the Majorana SYK model in the infinite temperature limit, $\beta=0$.
Here we will perform a similar calculation on the complex SYK model, and also obtain results at large $\beta J$. 
We define our renormalized OTOC by
\beq
\text{OTOC}=-\frac{\langle A(t)B(0)A(t)B(0)\rangle+\langle B(0)A(t)B(0)A(t)\rangle}{2\langle A A\rangle \langle B B\rangle}.
\eeq

\begin{figure}[!htb]
\center
\includegraphics[width=4.75in]{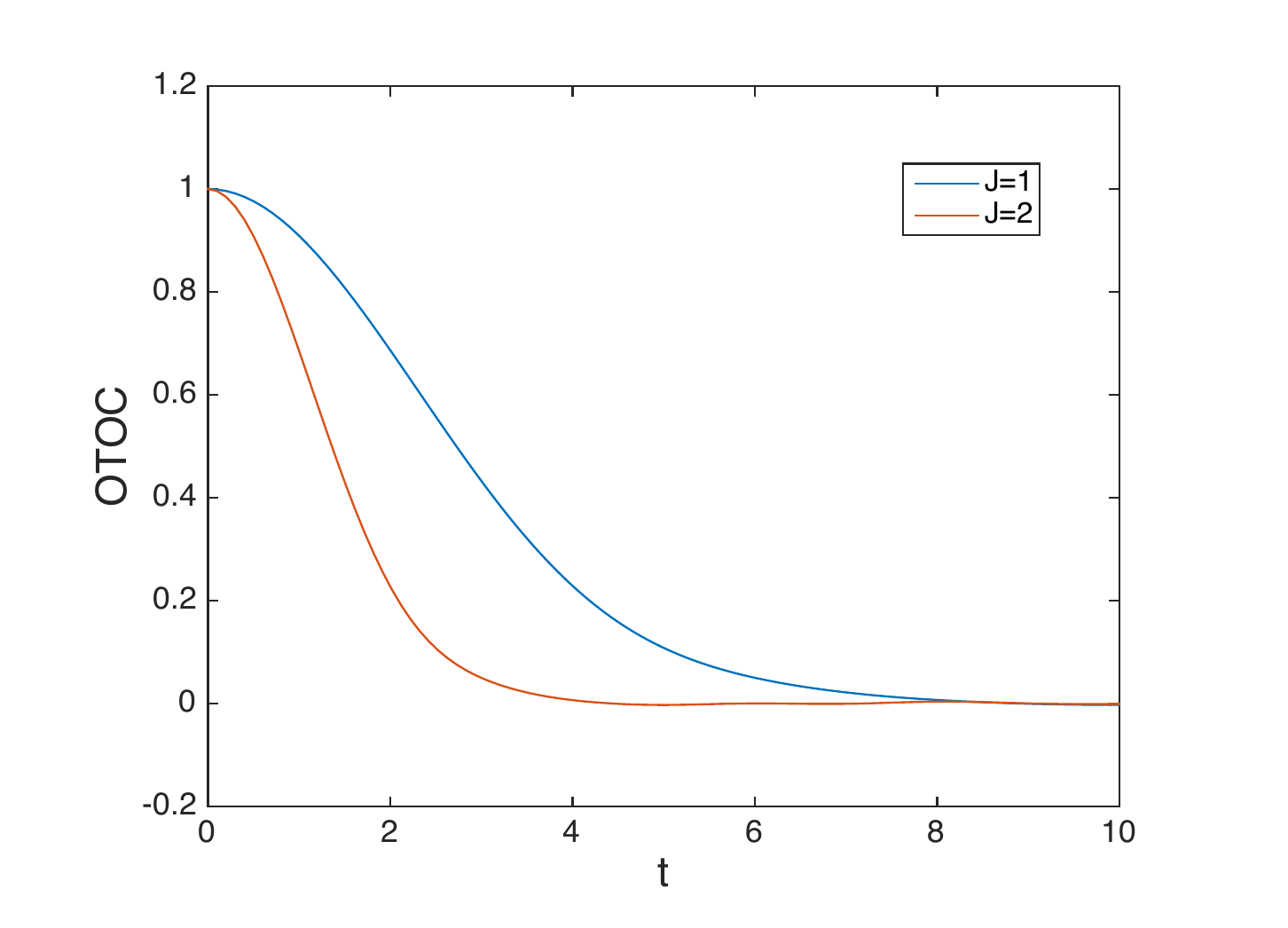}
\caption{OTOC as a function of time at infinite temperature with different interaction strength $J=1$ and $J=2$. Here the total system size $N=7$.}. 
\label{infiniteOTOC}
\end{figure}

\begin{figure}[!htb]
\center
\includegraphics[width=4.75in]{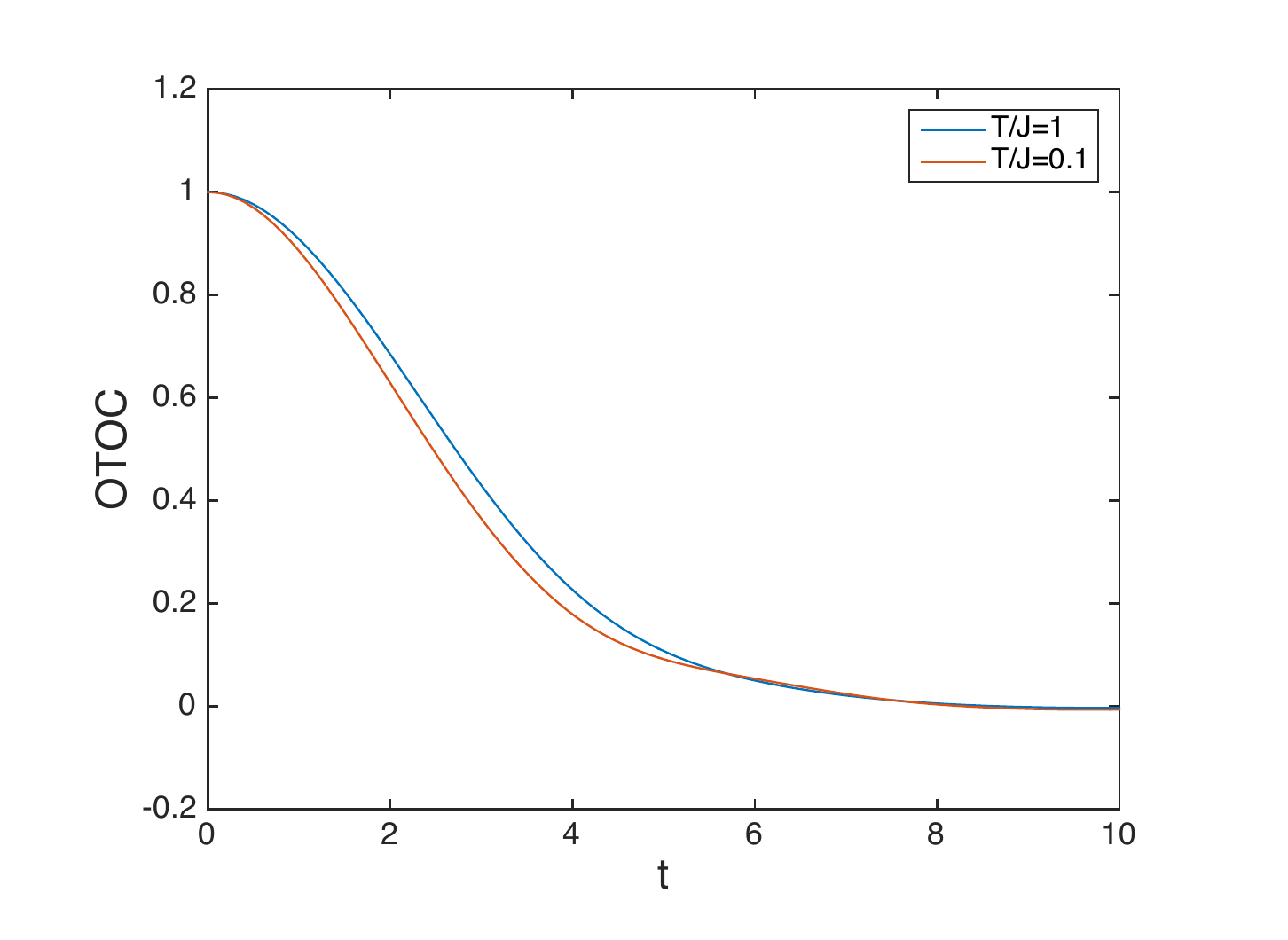}
\caption{OTOC as a function of time at different temperature with interaction strength $J=1$. Here the total system size $N=7$.}. 
\label{finiteOTOC}
\end{figure}

We choose the Hermitian Majorana operators $A=c_1+c_1^\dagger$, $B=c_2+c_2^\dagger$. The negative sign gives a positive initial value for OTOC. At infinite temperature, the result is shown in Fig.~\ref{infiniteOTOC}. We observe the fast scrambling effect from the quick decay of OTOC, and the early time decay rate $\lambda_L$ is proportional to $J$ as expected. Similar behavior is found in the Majorana SYK model \cite{hosur2015chaos}.
At finite temperature, although we can perform the computation in the strong coupling limit $\beta J\gg 1$, 
because of finite size effects, we do not get the predicted decay rate $\lambda_L=2\pi/\beta$. 
And the OTOC only has a weak dependence on $\beta$ even in the strong coupling limit as shown in Fig.~\ref{finiteOTOC}.
Theoretically \cite{maldacena2015abound}, in the large $N$ and strong coupling limit, $1-\text{OTOC}\sim ({\beta J}/{N}) e^{(2\pi/\beta) t}$. 
Fig.~\ref{finiteOTOC} does not display a large change in the exponent, and the pre-factor difference is also small. It is clearly that 
of our small system sizes, $J$ is the most relevant energy scale that controlling the chaos. 


\FloatBarrier

\section{SYK model for bosons}
\label{sec:EDb}

Now we consider a `cousin' of the present model: SYK model for hardcore bosons. 
The bosonic case was also considered in the early work \cite{sachdev1992gapless,PG98,GPS99,georges2001quantum} but with a large number
of bosons on each site. It was found that over most of the parameter regime the ground state had spin glass order. We will find evidence of similar behavior here.

\begin{figure}[!htb]
\center
\includegraphics[width=4.75in]{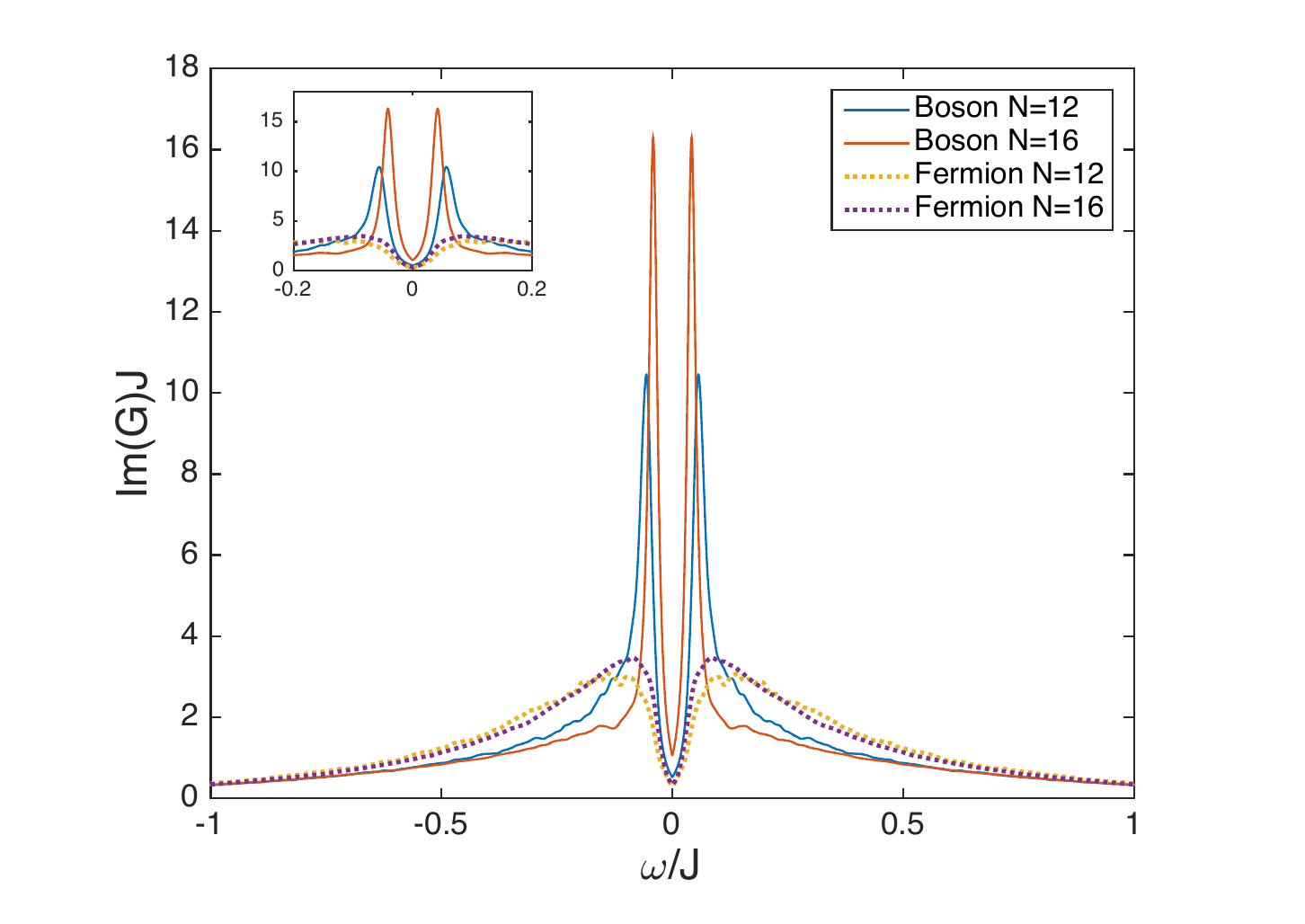}
\caption{Imaginary part of Green's function for hardcore boson and fermion model. The peak near the center gets much higher in the boson model when system size gets larger. The inset figure is zoomed in near $\omega=0$.}\label{BosonG}
\end{figure}

The Hamiltonian will be quite similar as Eq.~(\ref{H}), except that  because of the Bose statistics now the coefficients obey
\beq
J_{ji;kl} =  J_{ij;kl} \quad , \quad
J_{ij;lk} =  J_{ij;kl} \quad , \quad
J_{kl;ij} = J_{ij;kl}^\ast
\eeq
Hardcore boson satisfies $\left[b_i,b_j\right]=0$ for $i\neq j$ and $\{b_i,b_i^\dagger\}=1$. Also to make particle-hole symmetry (\ref{PH}) hold, 
we only consider pair hoping between different sites, \textit{i. e. \/} site indices $i,j,k,l$ are all different, 
and we drop the normal order correction terms. The spin formalism in ED will be even simpler, as we do not need to attach a Jordan-Wigner string of $\sigma_z$:
\beq
b_i=\sigma_i^-,\quad b_i^\dagger=\sigma_i^+
\eeq
We can define a similar Green's function for bosons:
\beq
G_B(t)=-i\theta(t)\langle\{b(t),b^\dagger(0)\}\rangle
\eeq
We identify the infinite time limit of $G_B$ as the Edward-Anderson order parameter $q_{EA}$, 
which can characterize long-time memory of spin-glass:
\beq
q_{EA}=\lim_{t\rightarrow\infty}G_B(t)
\eeq
Then $q_{EA}\neq0$ indicates that $G_B(\omega)\sim \delta(\omega)$. This is quite different from the fermionic case, where we have $G_F(z)\sim 1/\sqrt{z}$; this inverse square-root behavior also holds in the bosonic case without spin glass order \cite{sachdev1992gapless}.
Fig.~\ref{BosonG} is our result from ED, with a comparison between $G_B$ with $G_F$. 
It is evident that the behavior of $G_B$ is qualitatively different from $G_F$, and so an inverse square-root behavior is ruled out.
Instead, we can clearly see that, as system size gets larger, $G_B$'s peak value increases much faster than the $G_F$'s peak value. 
This supports the presence of spin glass order.

Unlike the fermionic case, $P^2=1$ for all $N$ in the bosonic model. We can apply similar symmetry argument as in Ref.~\cite{YLX16}: for the 
half-filled sector (only in even N cases), the level statistics obeys the Wigner-Dyson distribution of Gaussian orthogonal random matrix ensembles, while in other filling sectors, it obeys distribution of Gaussian unitary random matrix ensembles.

Our thermal entropy results for bosons are similar to the fermionic results: although the entropy eventually approaches 0 at zero temperature, 
there is still a trend of a larger low temperature entropy residue as the system size gets larger.

We have also computed the entanglement entropy for the ground state of the hardcore boson SYK model. It still satisfies volume law, and the entanglement entropy density is still quite close to $\ln 2$. Finally results for the OTOC are qualitatively similar to the fermionic results.

\FloatBarrier

\section{Conclusions}
\label{sec:conc}

We have presented exact diagonalization results on the fermionic SYK model. The trends in the computed Green's functions, high temperature
expansion, entropy density, and entanglement entropy all support the conclusion that the large $N$ limit approaches the compressible non-Fermi 
liquid state obtained in the earlier $N=\infty$ analysis. Note that the entropy density approaches a non-zero value in the limit $T \rightarrow 0$
taken {\em after\/} the $N \rightarrow \infty$, and so the ground state itself exhibits eigenstate thermalization. This conclusion
is also supported by the volume-law behavior of the entanglement entropy. 
The original model of Ref.~\onlinecite{sachdev1992gapless} was argued \cite{georges2001quantum} to have an instability to spin glass order 
at temperatures exponentially small in $\sqrt{M}$; the consonance between large $N$ theory and our finite $N$ numerics indicates
that the model in (\ref{H}) (with a random interaction with 4 indices \cite{kitaev2015talk}) does not have such an instability.

For the SYK model for hard-core bosons, our results for the single-particle Green's function were very different, and indicate the presence of 
spin glass order. Similar quantum spin glass states were examined in random models of bosons in Refs.~\onlinecite{GPS99,georges2001quantum}.

\section*{Acknowledgments} 
We thank Kartiek Agarwal, Shiang Fang, Tarun Grover, Yingfei Gu, Steve Gubser, Steve Shenker, and Yi-Zhuang You for valuable discussions. 
This research was supported by the NSF under Grant DMR-1360789
and  and MURI grant W911NF-14- 1-0003 from ARO.
Research at Perimeter Institute is supported by the Government of Canada through Industry Canada 
and by the Province of Ontario through the Ministry of Research and Innovation.

\bibliography{ref.bib}

\begin{thebibliography}{18}%
\makeatletter
\providecommand \@ifxundefined [1]{%
 \@ifx{#1\undefined}
}%
\providecommand \@ifnum [1]{%
 \ifnum #1\expandafter \@firstoftwo
 \else \expandafter \@secondoftwo
 \fi
}%
\providecommand \@ifx [1]{%
 \ifx #1\expandafter \@firstoftwo
 \else \expandafter \@secondoftwo
 \fi
}%
\providecommand \natexlab [1]{#1}%
\providecommand \enquote  [1]{``#1''}%
\providecommand \bibnamefont  [1]{#1}%
\providecommand \bibfnamefont [1]{#1}%
\providecommand \citenamefont [1]{#1}%
\providecommand \href@noop [0]{\@secondoftwo}%
\providecommand \href [0]{\begingroup \@sanitize@url \@href}%
\providecommand \@href[1]{\@@startlink{#1}\@@href}%
\providecommand \@@href[1]{\endgroup#1\@@endlink}%
\providecommand \@sanitize@url [0]{\catcode `\\12\catcode `\$12\catcode
  `\&12\catcode `\#12\catcode `\^12\catcode `\_12\catcode `\%12\relax}%
\providecommand \@@startlink[1]{}%
\providecommand \@@endlink[0]{}%
\providecommand \url  [0]{\begingroup\@sanitize@url \@url }%
\providecommand \@url [1]{\endgroup\@href {#1}{\urlprefix }}%
\providecommand \urlprefix  [0]{URL }%
\providecommand \Eprint [0]{\href }%
\providecommand \doibase [0]{http://dx.doi.org/}%
\providecommand \selectlanguage [0]{\@gobble}%
\providecommand \bibinfo  [0]{\@secondoftwo}%
\providecommand \bibfield  [0]{\@secondoftwo}%
\providecommand \translation [1]{[#1]}%
\providecommand \BibitemOpen [0]{}%
\providecommand \bibitemStop [0]{}%
\providecommand \bibitemNoStop [0]{.\EOS\space}%
\providecommand \EOS [0]{\spacefactor3000\relax}%
\providecommand \BibitemShut  [1]{\csname bibitem#1\endcsname}%
\let\auto@bib@innerbib\@empty
\bibitem [{\citenamefont {{Sachdev}}\ and\ \citenamefont
  {{Ye}}(1993)}]{sachdev1992gapless}%
  \BibitemOpen
  \bibfield  {author} {\bibinfo {author} {\bibfnamefont {S.}~\bibnamefont
  {{Sachdev}}}\ and\ \bibinfo {author} {\bibfnamefont {J.}~\bibnamefont
  {{Ye}}},\ }\bibfield  {title} {\enquote {\bibinfo {title} {{Gapless
  spin-fluid ground state in a random quantum Heisenberg magnet}},}\ }\href
  {\doibase 10.1103/PhysRevLett.70.3339} {\bibfield  {journal} {\bibinfo
  {journal} {\prl}\ }\textbf {\bibinfo {volume} {70}},\ \bibinfo {pages} {3339}
  (\bibinfo {year} {1993})},\ \Eprint {http://arxiv.org/abs/cond-mat/9212030}
  {cond-mat/9212030} \BibitemShut {NoStop}%
\bibitem [{\citenamefont {{Parcollet}}\ and\ \citenamefont
  {{Georges}}(1999)}]{PG98}%
  \BibitemOpen
  \bibfield  {author} {\bibinfo {author} {\bibfnamefont {O.}~\bibnamefont
  {{Parcollet}}}\ and\ \bibinfo {author} {\bibfnamefont {A.}~\bibnamefont
  {{Georges}}},\ }\bibfield  {title} {\enquote {\bibinfo {title}
  {{Non-Fermi-liquid regime of a doped Mott insulator}},}\ }\href {\doibase
  10.1103/PhysRevB.59.5341} {\bibfield  {journal} {\bibinfo  {journal} {\prb}\
  }\textbf {\bibinfo {volume} {59}},\ \bibinfo {pages} {5341} (\bibinfo {year}
  {1999})},\ \Eprint {http://arxiv.org/abs/cond-mat/9806119} {cond-mat/9806119}
  \BibitemShut {NoStop}%
\bibitem [{\citenamefont {{Georges}}\ \emph {et~al.}(2000)\citenamefont
  {{Georges}}, \citenamefont {{Parcollet}},\ and\ \citenamefont
  {{Sachdev}}}]{GPS99}%
  \BibitemOpen
  \bibfield  {author} {\bibinfo {author} {\bibfnamefont {A.}~\bibnamefont
  {{Georges}}}, \bibinfo {author} {\bibfnamefont {O.}~\bibnamefont
  {{Parcollet}}}, \ and\ \bibinfo {author} {\bibfnamefont {S.}~\bibnamefont
  {{Sachdev}}},\ }\bibfield  {title} {\enquote {\bibinfo {title} {{Mean Field
  Theory of a Quantum Heisenberg Spin Glass}},}\ }\href {\doibase
  10.1103/PhysRevLett.85.840} {\bibfield  {journal} {\bibinfo  {journal}
  {\prl}\ }\textbf {\bibinfo {volume} {85}},\ \bibinfo {pages} {840} (\bibinfo
  {year} {2000})},\ \Eprint {http://arxiv.org/abs/cond-mat/9909239}
  {cond-mat/9909239} \BibitemShut {NoStop}%
\bibitem [{\citenamefont {{Georges}}\ \emph {et~al.}(2001)\citenamefont
  {{Georges}}, \citenamefont {{Parcollet}},\ and\ \citenamefont
  {{Sachdev}}}]{georges2001quantum}%
  \BibitemOpen
  \bibfield  {author} {\bibinfo {author} {\bibfnamefont {A.}~\bibnamefont
  {{Georges}}}, \bibinfo {author} {\bibfnamefont {O.}~\bibnamefont
  {{Parcollet}}}, \ and\ \bibinfo {author} {\bibfnamefont {S.}~\bibnamefont
  {{Sachdev}}},\ }\bibfield  {title} {\enquote {\bibinfo {title} {{Quantum
  fluctuations of a nearly critical Heisenberg spin glass}},}\ }\href {\doibase
  10.1103/PhysRevB.63.134406} {\bibfield  {journal} {\bibinfo  {journal}
  {\prb}\ }\textbf {\bibinfo {volume} {63}},\ \bibinfo {eid} {134406} (\bibinfo
  {year} {2001})},\ \Eprint {http://arxiv.org/abs/cond-mat/0009388}
  {cond-mat/0009388} \BibitemShut {NoStop}%
\bibitem [{\citenamefont {{Arrachea}}\ and\ \citenamefont
  {{Rozenberg}}(2002)}]{MJR02}%
  \BibitemOpen
  \bibfield  {author} {\bibinfo {author} {\bibfnamefont {L.}~\bibnamefont
  {{Arrachea}}}\ and\ \bibinfo {author} {\bibfnamefont {M.~J.}\ \bibnamefont
  {{Rozenberg}}},\ }\bibfield  {title} {\enquote {\bibinfo {title}
  {{Infinite-range quantum random Heisenberg magnet}},}\ }\href {\doibase
  10.1103/PhysRevB.65.224430} {\bibfield  {journal} {\bibinfo  {journal}
  {\prb}\ }\textbf {\bibinfo {volume} {65}},\ \bibinfo {eid} {224430} (\bibinfo
  {year} {2002})},\ \Eprint {http://arxiv.org/abs/cond-mat/0203537}
  {cond-mat/0203537} \BibitemShut {NoStop}%
\bibitem [{\citenamefont {{Camjayi}}\ and\ \citenamefont
  {{Rozenberg}}(2003)}]{MJR03}%
  \BibitemOpen
  \bibfield  {author} {\bibinfo {author} {\bibfnamefont {A.}~\bibnamefont
  {{Camjayi}}}\ and\ \bibinfo {author} {\bibfnamefont {M.~J.}\ \bibnamefont
  {{Rozenberg}}},\ }\bibfield  {title} {\enquote {\bibinfo {title} {{Quantum
  and Thermal Fluctuations in the SU($N$) Heisenberg Spin-Glass Model near the
  Quantum Critical Point}},}\ }\href {\doibase 10.1103/PhysRevLett.90.217202}
  {\bibfield  {journal} {\bibinfo  {journal} {\prl}\ }\textbf {\bibinfo
  {volume} {90}},\ \bibinfo {eid} {217202} (\bibinfo {year} {2003})},\ \Eprint
  {http://arxiv.org/abs/cond-mat/0210407} {cond-mat/0210407} \BibitemShut
  {NoStop}%
\bibitem [{\citenamefont {Sachdev}(2010{\natexlab{a}})}]{SS10}%
  \BibitemOpen
  \bibfield  {author} {\bibinfo {author} {\bibfnamefont {S.}~\bibnamefont
  {Sachdev}},\ }\bibfield  {title} {\enquote {\bibinfo {title} {{Holographic
  metals and the fractionalized Fermi liquid}},}\ }\href {\doibase
  10.1103/PhysRevLett.105.151602} {\bibfield  {journal} {\bibinfo  {journal}
  {Phys. Rev. Lett.}\ }\textbf {\bibinfo {volume} {105}},\ \bibinfo {pages}
  {151602} (\bibinfo {year} {2010}{\natexlab{a}})},\ \Eprint
  {http://arxiv.org/abs/1006.3794} {arXiv:1006.3794 [hep-th]} \BibitemShut
  {NoStop}%
\bibitem [{\citenamefont {Sachdev}(2010{\natexlab{b}})}]{SS10b}%
  \BibitemOpen
  \bibfield  {author} {\bibinfo {author} {\bibfnamefont {S.}~\bibnamefont
  {Sachdev}},\ }\bibfield  {title} {\enquote {\bibinfo {title} {{Strange metals
  and the AdS/CFT correspondence}},}\ }\href {\doibase
  10.1088/1742-5468/2010/11/P11022} {\bibfield  {journal} {\bibinfo  {journal}
  {J. Stat. Mech.}\ }\textbf {\bibinfo {volume} {1011}},\ \bibinfo {pages}
  {P11022} (\bibinfo {year} {2010}{\natexlab{b}})},\ \Eprint
  {http://arxiv.org/abs/1010.0682} {arXiv:1010.0682 [cond-mat.str-el]}
  \BibitemShut {NoStop}%
\bibitem [{\citenamefont {{Kitaev}}(2015)}]{kitaev2015talk}%
  \BibitemOpen
  \bibfield  {author} {\bibinfo {author} {\bibfnamefont {A.~Y.}\ \bibnamefont
  {{Kitaev}}},\ }\bibfield  {title} {\enquote {\bibinfo {title} {{Talks at
  KITP, University of California, Santa Barbara}},}\ }\href
  {http://online.kitp.ucsb.edu/online/entangled15/} {\bibfield  {journal}
  {\bibinfo  {journal} {Entanglement in Strongly-Correlated Quantum Matter}\ }
  (\bibinfo {year} {2015})}\BibitemShut {NoStop}%
\bibitem [{\citenamefont {{Sachdev}}(2015)}]{sachdev2015Bh}%
  \BibitemOpen
  \bibfield  {author} {\bibinfo {author} {\bibfnamefont {S.}~\bibnamefont
  {{Sachdev}}},\ }\bibfield  {title} {\enquote {\bibinfo {title}
  {{Bekenstein-Hawking Entropy and Strange Metals}},}\ }\href {\doibase
  10.1103/PhysRevX.5.041025} {\bibfield  {journal} {\bibinfo  {journal} {Phys.
  Rev. X}\ }\textbf {\bibinfo {volume} {5}},\ \bibinfo {eid} {041025} (\bibinfo
  {year} {2015})},\ \Eprint {http://arxiv.org/abs/1506.05111} {arXiv:1506.05111
  [hep-th]} \BibitemShut {NoStop}%
\bibitem [{\citenamefont {{Maldacena}}\ \emph {et~al.}(2015)\citenamefont
  {{Maldacena}}, \citenamefont {{Shenker}},\ and\ \citenamefont
  {{Stanford}}}]{maldacena2015abound}%
  \BibitemOpen
  \bibfield  {author} {\bibinfo {author} {\bibfnamefont {J.}~\bibnamefont
  {{Maldacena}}}, \bibinfo {author} {\bibfnamefont {S.~H.}\ \bibnamefont
  {{Shenker}}}, \ and\ \bibinfo {author} {\bibfnamefont {D.}~\bibnamefont
  {{Stanford}}},\ }\bibfield  {title} {\enquote {\bibinfo {title} {{A bound on
  chaos}},}\ }\href@noop {} {\bibfield  {journal} {\bibinfo  {journal} {ArXiv
  e-prints}\ } (\bibinfo {year} {2015})},\ \Eprint
  {http://arxiv.org/abs/1503.01409} {arXiv:1503.01409 [hep-th]} \BibitemShut
  {NoStop}%
\bibitem [{\citenamefont {Hosur}\ \emph {et~al.}(2016)\citenamefont {Hosur},
  \citenamefont {Qi}, \citenamefont {Roberts},\ and\ \citenamefont
  {Yoshida}}]{hosur2015chaos}%
  \BibitemOpen
  \bibfield  {author} {\bibinfo {author} {\bibfnamefont {P.}~\bibnamefont
  {Hosur}}, \bibinfo {author} {\bibfnamefont {X.-L.}\ \bibnamefont {Qi}},
  \bibinfo {author} {\bibfnamefont {D.~A.}\ \bibnamefont {Roberts}}, \ and\
  \bibinfo {author} {\bibfnamefont {B.}~\bibnamefont {Yoshida}},\ }\bibfield
  {title} {\enquote {\bibinfo {title} {{Chaos in quantum channels}},}\ }\href
  {\doibase 10.1007/JHEP02(2016)004} {\bibfield  {journal} {\bibinfo  {journal}
  {JHEP}\ }\textbf {\bibinfo {volume} {02}},\ \bibinfo {pages} {004} (\bibinfo
  {year} {2016})},\ \Eprint {http://arxiv.org/abs/1511.04021} {arXiv:1511.04021
  [hep-th]} \BibitemShut {NoStop}%
\bibitem [{\citenamefont {{Polchinski}}\ and\ \citenamefont
  {{Rosenhaus}}(2016)}]{JPRV16}%
  \BibitemOpen
  \bibfield  {author} {\bibinfo {author} {\bibfnamefont {J.}~\bibnamefont
  {{Polchinski}}}\ and\ \bibinfo {author} {\bibfnamefont {V.}~\bibnamefont
  {{Rosenhaus}}},\ }\bibfield  {title} {\enquote {\bibinfo {title} {{The
  Spectrum in the Sachdev-Ye-Kitaev Model}},}\ }\href@noop {} {\bibfield
  {journal} {\bibinfo  {journal} {ArXiv e-prints}\ } (\bibinfo {year}
  {2016})},\ \Eprint {http://arxiv.org/abs/1601.06768} {arXiv:1601.06768
  [hep-th]} \BibitemShut {NoStop}%
\bibitem [{\citenamefont {{You}}\ \emph {et~al.}(2016)\citenamefont {{You}},
  \citenamefont {{Ludwig}},\ and\ \citenamefont {{Xu}}}]{YLX16}%
  \BibitemOpen
  \bibfield  {author} {\bibinfo {author} {\bibfnamefont {Y.-Z.}\ \bibnamefont
  {{You}}}, \bibinfo {author} {\bibfnamefont {A.~W.~W.}\ \bibnamefont
  {{Ludwig}}}, \ and\ \bibinfo {author} {\bibfnamefont {C.}~\bibnamefont
  {{Xu}}},\ }\bibfield  {title} {\enquote {\bibinfo {title} {{Sachdev-Ye-Kitaev
  Model and Thermalization on the Boundary of Many-Body Localized Fermionic
  Symmetry Protected Topological States}},}\ }\href@noop {} {\bibfield
  {journal} {\bibinfo  {journal} {ArXiv e-prints}\ } (\bibinfo {year}
  {2016})},\ \Eprint {http://arxiv.org/abs/1602.06964} {arXiv:1602.06964
  [cond-mat.str-el]} \BibitemShut {NoStop}%
\bibitem [{\citenamefont {{Anninos}}\ \emph {et~al.}(2016)\citenamefont
  {{Anninos}}, \citenamefont {{Anous}},\ and\ \citenamefont
  {{Denef}}}]{Denef16}%
  \BibitemOpen
  \bibfield  {author} {\bibinfo {author} {\bibfnamefont {D.}~\bibnamefont
  {{Anninos}}}, \bibinfo {author} {\bibfnamefont {T.}~\bibnamefont {{Anous}}},
  \ and\ \bibinfo {author} {\bibfnamefont {F.}~\bibnamefont {{Denef}}},\
  }\bibfield  {title} {\enquote {\bibinfo {title} {{Disordered Quivers and Cold
  Horizons}},}\ }\href@noop {} {\bibfield  {journal} {\bibinfo  {journal}
  {ArXiv e-prints}\ } (\bibinfo {year} {2016})},\ \Eprint
  {http://arxiv.org/abs/1603.00453} {arXiv:1603.00453 [hep-th]} \BibitemShut
  {NoStop}%
\bibitem [{\citenamefont {{Jevicki}}\ \emph {et~al.}(2016)\citenamefont
  {{Jevicki}}, \citenamefont {{Suzuki}},\ and\ \citenamefont
  {{Yoon}}}]{Jevicki16}%
  \BibitemOpen
  \bibfield  {author} {\bibinfo {author} {\bibfnamefont {A.}~\bibnamefont
  {{Jevicki}}}, \bibinfo {author} {\bibfnamefont {K.}~\bibnamefont {{Suzuki}}},
  \ and\ \bibinfo {author} {\bibfnamefont {J.}~\bibnamefont {{Yoon}}},\
  }\bibfield  {title} {\enquote {\bibinfo {title} {{Bi-Local Holography in the
  SYK Model}},}\ }\href@noop {} {\bibfield  {journal} {\bibinfo  {journal}
  {ArXiv e-prints}\ } (\bibinfo {year} {2016})},\ \Eprint
  {http://arxiv.org/abs/1603.06246} {arXiv:1603.06246 [hep-th]} \BibitemShut
  {NoStop}%
\bibitem [{\citenamefont {{DeWolfe}}\ \emph {et~al.}(2012)\citenamefont
  {{DeWolfe}}, \citenamefont {{Gubser}},\ and\ \citenamefont
  {{Rosen}}}]{DGR11}%
  \BibitemOpen
  \bibfield  {author} {\bibinfo {author} {\bibfnamefont {O.}~\bibnamefont
  {{DeWolfe}}}, \bibinfo {author} {\bibfnamefont {S.~S.}\ \bibnamefont
  {{Gubser}}}, \ and\ \bibinfo {author} {\bibfnamefont {C.}~\bibnamefont
  {{Rosen}}},\ }\bibfield  {title} {\enquote {\bibinfo {title} {{Fermi Surfaces
  in Maximal Gauged Supergravity}},}\ }\href {\doibase
  10.1103/PhysRevLett.108.251601} {\bibfield  {journal} {\bibinfo  {journal}
  {Phys. Rev. Lett.}\ }\textbf {\bibinfo {volume} {108}},\ \bibinfo {eid}
  {251601} (\bibinfo {year} {2012})},\ \Eprint {http://arxiv.org/abs/1112.3036}
  {arXiv:1112.3036 [hep-th]} \BibitemShut {NoStop}%
\bibitem [{tar()}]{tarun}%
  \BibitemOpen
  \href@noop {} {\ }\bibinfo {note} {{We thank Tarun Grover and Yi-Zhuang You
  for pointing this out to us.}}\BibitemShut {Stop}%
\end{thebibliography}%
\end{document}